\documentclass{aa}
\usepackage{txfonts}
\usepackage{graphicx}
\usepackage{natbib}
\bibpunct{(}{)}{;}{a}{}{,}

\begin{document}

\newcommand{\eps}{\varepsilon}
\newcommand{\crp}{\mathrm{CRp}}
\newcommand{\cre}{\mathrm{CRe}}
\newcommand{\e}{\mathrm{e}}
\newcommand{\p}{\mathrm{p}}
\newcommand{\gas}{\mathrm{gas}}
\newcommand{\dd}{\mathrm{d}}
\newcommand{\B}{\mathcal{B}}
\newcommand{\I}{\mathcal{I}}
\newcommand{\rmn}{\mathrm}
\newcommand{\bld}{\mathbf}
\newcommand{\dps}{\displaystyle}
\newcommand{\bra}{\langle}
\newcommand{\ket}{\rangle}

\title{Unveiling the composition of radio plasma bubbles in galaxy clusters
  with the Sunyaev-Zel'dovich effect}
\titlerunning{Unveiling the composition of radio plasma bubbles}

\author{C. Pfrommer \inst{1}, 
        T. A. En{\ss}lin \inst{1}, 
        \and C. L. Sarazin \inst{2}}
\authorrunning{C. Pfrommer, T. A. En{\ss}lin, \& C. L. Sarazin}

\institute{Max-Planck-Institut f\"{u}r Astrophysik,
Karl-Schwarzschild-Str.1, PO Box 1317, 85741 Garching, Germany
\and
Department of Astronomy, University of Virginia, PO Box 3818, 
Charlottesville, VA 22903-0818, USA} 

\offprints{Christoph Pfrommer, \\ \email{pfrommer@mpa-garching.mpg.de}}
\date{\today}

\abstract{The {\em Chandra} X-ray Observatory is finding a large number of
  cavities in the X-ray emitting intra-cluster medium which often coincide with
  the lobes of the central radio galaxy. We propose high-resolution
  Sunyaev-Zel'dovich (SZ) observations in order to infer the still unknown
  dynamically dominating component of the radio plasma bubbles. This work
  calculates the thermal and relativistic SZ emission of different compositions
  of these plasma bubbles while simultaneously allowing for the cluster's
  kinetic SZ effect.  As examples, we present simulations of an {\em Atacama
    Large Millimeter Array (ALMA)} observation and of a {\em Green Bank
    Telescope (GBT)} observation of the cores of the Perseus cluster and
  Abell~2052.  We predict a 5$\sigma$ detection of the southern radio bubble of
  Perseus in a few hours with the {\it GBT} and {\it ALMA} while assuming a
  relativistic electron population within the bubble.  In Abell~2052, a similar
  detection would require a few ten hours with either telescope, the longer
  exposures mainly being the result of the higher redshift and the lower
  central temperature of this cluster.  Future high-sensitivity multi-frequency
  SZ observations will be able to infer the energy spectrum of the dynamically
  dominating electron population in order to measure its temperature or
  spectral characteristics.  This knowledge can yield indirect indications for
  an underlying radio jet model.
}
\maketitle

\section{Introduction}
\label{sec:intro}
The {\em Chandra} X-ray Observatory is detecting numerous X-ray cavities in
clusters of galaxies confirming pioneering detections of the {\em ROSAT}
satellite. Prominent examples include the \object{Perseus cluster}
\citep{1993MNRAS.264L..25B, 2000MNRAS.318L..65F}, the \object{Cygnus-A cluster}
\citep{1994MNRAS.270..173C}, the \object{Hydra-A cluster}
\citep{2000ApJ...534L.135M}, \object{Abell 2597} \citep{2001ApJ...562L.149M},
\object{Abell 4059} \citep{1998ApJ...496..728H, 2002ApJ...569L..79H},
\object{Abell 2199} \citep{fabian2001moriond}, \object{Abell 2052}
\citep{2001ApJ...558L..15B}, the vicinity of \object{M84} in the \object{Virgo
  cluster} \citep{2001ApJ...547L.107F}, the \object{RBS797 cluster}
\citep{2001A&A...376L..27S}, and the \object{MKW3s cluster}
\citep{2002ApJ...567L..37M}.  They are produced by the release of radio plasma
from active galactic nuclei (AGN) which are typically hosted by a cD galaxy
located at the cluster center and mainly reside within the cool cores of galaxy
clusters.  

While radio synchrotron emission provides evidence for the existence of cosmic
ray electrons (CRes) and magnetic fields, the detailed composition of the
plasma bubble governing its dynamics is still unknown.  Minimum energy or
equipartition estimates of the nonthermal pressure in the radio bubbles give
values which are typically a factor of ten smaller than the pressures required
to inflate and maintain the bubbles as determined from the surrounding X-ray
gas \citep[e.g.,][]{2001ApJ...558L..15B}.  This indicates that the standard
minimum energy or equipartition radio arguments are missing the main component
of the pressure and energy content of the radio lobes.  Possibilities include
magnetic fields, cosmic ray proton (CRp) or CRe power-law distributions, or
very hot thermal gas.  Solving this enigma would yield further insight into
physical processes within cool cores \citep{2003MNRAS.343..719D} as well as
provide hints about the composition of relativistic outflows of radio galaxies
because plasma bubbles represent the relic fluid of jets
\citep[e.g.][]{1998MNRAS.293..288C, 1998tx19.confE.407H, 2000ApJ...534..109S}.

Additionally, some of the clusters exhibit cavities in the X-ray emitting
intra-cluster medium (ICM) without detectable high frequency radio emission,
for instance in \object{Perseus}, \object{Abell 2597}, \object{Abell 4059}, and
the \object{MKW3s cluster}. This category of X-ray cavities is also believed to
be filled with radio plasma, but during the buoyant rise of the light radio
plasma bubble in the cluster's potential
\citep{gull1973,2000A&A...356..788C,2001ApJ...554..261C,2001MNRAS.325..676B}
the resulting adiabatic expansion and synchrotron/inverse Compton losses
dwindles the observable radio emitting electron population producing a
so-called {\em ghost cavity} or {\em radio ghost} \citep{1999dtrp.conf..275E}.
Possible entrainment of the ICM into the plasma bubble and subsequent Coulomb
heating by CRes generates further uncertainty of the composition of the ghost
cavity.

If the radio bubbles contain a significant amount of very hot thermal gas, this
might be detected by X-ray observations.  However, this is quite difficult
\citep[e.g.,][]{2003ApJ...585..227B} due to the projected foreground and
background cluster emission, and the fact that the X-ray emissivity is
proportional to the square of the density.  If most of the pressure in the
radio bubbles were due to the very hot, low density thermal gas, it would have
a very low X-ray emissivity. Observationally, there has been a claim by
\citet{2002ApJ...567L..37M} to have seen hot X-ray emitting gas within the
ghost cavity of the MKW3s cluster.  Obviously, it might be more useful to
observe the radio bubbles with a technique which was sensitive to thermal gas
pressure, rather than density squared, as pressure is the quantity which is
missing.  For this reason, in this paper we propose high resolution
Sunyaev-Zel'dovich (SZ) radio observations of the radio bubbles and radio
ghosts in clusters, as the thermal SZ effect directly measures the thermal
electron pressure in the gas. The thermal SZ effect arises because photons of
the cosmic microwave background (CMB) experience inverse Compton collisions
with thermal electrons of the hot plasma inside clusters of galaxies and are
spectrally redistributed \citep[e.g.][]{1972SZorig, 1980ARA&A..18..537S,
  1995ARA&A..33..541R}. The proposed measurement is able to infer directly the
composition of radio plasma bubbles and radio ghosts while indirectly obtaining
indications for a specific underlying jet model.

The article is organized as follows: After basic definitions concerning the
thermal, kinetic, and relativistic SZ effect in Sect.~\ref{sec:rSZ}, we
introduce a toy model in Sect.~\ref{sec:toymodel} describing projected maps of
the SZ flux decrement with spherically symmetric radio plasma bubbles.  The
models for the cool core regions of the Perseus cluster and Abell~2052 are
described in Sect.~\ref{sec:perseus}.  Simulating an {\em Atacama Large
  Millimeter Array (ALMA)} and a {\em Green Bank Telescope (GBT)} observation
of both clusters in Sect.~\ref{sec:synthetic}, we examine whether the plasma
bubbles are detectable by the SZ flux decrement.  Five physically different
scenarios for the plasma composition of the bubbles are investigated
exemplarily using three characteristic SZ frequencies in
Sects.~\ref{sec:composition} and \ref{sec:kinSZ}. Finally, we discuss observing
strategies for {\em ALMA} and {\em GBT}.  Throughout the paper, we assume a
$\Lambda$CDM cosmology and the Hubble parameter at the present time of $H_{0} =
70\, h_\rmn{70}~{\rm km~s^{-1}~Mpc^{-1}}$.

\section{Sunyaev-Zel'dovich effect}
\label{sec:rSZ}

The SZ effect arises because CMB photons experience inverse Compton (IC)
scattering off electrons of the dilute intra-cluster plasma \citep[for a
comprehensive review, see][] {1999PhR...310...97B}.  At the angular position of
galaxy clusters, the CMB spectrum is modulated as photons are redistributed
from the low-frequency part of the spectrum below a characteristic crossover
frequency $\nu_\rmn{c}$ to higher frequencies. For non-relativistic electron
populations, $\nu_\rmn{c} \simeq 217\mbox{ GHz}$, while this characteristic
frequency shifts towards higher values for more relativistic scattering
electrons.

The relative change $\delta i(x)$ in flux density as a function of
dimensionless frequency $x=h\nu / (k T_\mathrm{CMB})$ for a line-of-sight
through a galaxy cluster is given by
\begin{eqnarray}
  \delta i(x) &=& 
  g(x)\, y_\rmn{gas}\, \left[ 1+\delta(x,T_\e) \right] - h(x) \, w_\gas 
  \nonumber\\
  \label{eq:deltai}
  && +\, [j(x) - i(x)]\, \tau_\rmn{rel}\,,
\end{eqnarray}
with the Planckian distribution function of the CMB
\begin{equation}
  \label{eq:Inu}
  I(x) = i_0 i(x) = i_0 \frac{x^3}{\e^x - 1},
\end{equation}
and $i_0 = 2(k T_\rmn{CMB})^3/(hc)^2$ where $T_\rmn{CMB} =
2.725\mbox{ K}$, $k$, $h$ and $c$ denote the average CMB temperature,
Boltzmann's constant, Planck's constant, and the speed of light, respectively.

The first term in Eqn.~(\ref{eq:deltai}) arises because of the thermal motion
of non-relativistic electrons (thermal SZ effect) and gives the spectral
distortion
\begin{equation}
  \label{eq:g}
  g(x) = \frac{x^4 \e^x}{\left(\e^x - 1\right)^2}
  \left(x\,\frac{\e^x + 1}{\e^x - 1} - 4\right).
\end{equation}
The amplitude of the thermal SZ effect is known as the thermal Comptonization
parameter $y_\rmn{gas}$ that is defined as the line-of-sight integration of the
temperature weighted thermal electron density from the observer to the last
scattering surface of the CMB at redshift $z = 1100$:
\begin{equation}
  \label{eq:ygas}
  y_\rmn{gas} \equiv
  \frac{\sigma_\mathrm{T}}{m_\e c^2}\int\dd l\,n_\rmn{e,gas}\, k T_\e.
\end{equation}
Here, $\sigma_\rmn{T}$ denotes the Thompson cross section, $m_\e$ the electron
rest mass, $T_\e$ and $n_\rmn{e,gas}$ are electron temperature and thermal
electron number density, respectively.  For non-relativistic electrons the
relativistic correction term is zero, $\delta(x,T_\e) = 0$, but for hot
clusters even the thermal electrons have relativistic corrections, which will
modify the thermal SZ effect \citep{1979ApJ...232..348W}.  These corrections
have been calculated in the literature \citep[see e.g.][]{1995ApJ...445...33R,
  2000A&A...360..417E, 2001ApJ...554...74D, 2004A&A...417..827I}, and can be
used to measure the cluster temperature purely from SZ
observations~\citep[e.g.][]{2002ApJ...573L..69H}.

The second term in Eqn.~(\ref{eq:deltai}) describes an additional spectral
distortion of the CMB spectrum due to the Doppler effect of the bulk motion of
the cluster itself relative to the rest frame of the CMB. If the component of
the cluster's peculiar velocity is projected along the line-of-sight, then the
Doppler effect leads to a spectral distortion referred to as the kinetic SZ
effect with the spectral signature
\begin{equation}
h(x) = \frac{x^4 \,e^x}{(e^x -1)^2}.
\end{equation}
The amplitude of the kinetic SZ effect depends on the kinetic Comptonization
parameter $w_\gas$ that is defined as 
\begin{equation}
\label{eq:barbeta}
w_\gas \equiv \bar{\beta}_\gas \, \tau_\gas = 
\sigma_{\rm T} \int \dd l\, n_{\e ,\gas}\,\bar{\beta}_\gas,
\end{equation}
where $\tau_{\rm gas}$ is the Thomson optical depth, ${\bar{\upsilon}}_{\rm
  gas}$ is the average line-of-sight streaming velocity of the thermal gas,
${\bar{\beta}}_{\rm gas} \equiv {\bar{\upsilon}}_{\rm gas} / c$, and
${\bar{\beta}}_{\rm gas} < 0$ if the gas is approaching the observer.

Finally, the third term in Eqn.~(\ref{eq:deltai}) takes account of Compton
scattering with relativistic electrons that exhibit an optical depth of
\begin{equation}
  \label{eq:tau}
  \tau_\rmn{rel} = \sigma_\rmn{T}\int\dd l\,n_\rmn{e,rel}.
\end{equation}
The flux scattered to other frequencies is $i(x)\tau_\rmn{rel}$ while
$j(x)\tau_\rmn{rel}$ is the flux scattered from other frequencies to $x=h\nu /
(k T_\mathrm{CMB})$.  It is worth noting, that in the limit of
ultra-relativistic electrons and for $x<10$, one can neglect the scattered
flux, because $j(x) \ll i(x)$.  In the following, we drop this approximation
and consider the general case.  The scattered flux can be expressed in terms of
the photon redistribution function for a mono-energetic electron distribution
$P(t;p)$, where the frequency of a scattered photon is shifted by a factor $t$:
\begin{equation}
\label{eq:j} 
j(x) = \int_0^\infty\!  \dd t \, \int_0^\infty\!  \dd p \, 
f_\e (p)\, P(t; p)\, i(x/t).
\end{equation}
For a given electron spectrum $f_\e(p) \,\dd p$ with the normalized electron
momentum $p = \beta_\e \gamma_\e$ and $\int \dd p \, f_\e (p) =1$, this
redistribution function can be derived following the kinematic considerations
of \citet{1979ApJ...232..348W} of the IC scattering in the Thomson regime,
where $\gamma_{\rm e} \,h\nu \ll m_{\rm e}c^2$ is valid.  We use the compact
formula for the photon redistribution function which was derived by
\citet{2000A&A...360..417E}:
\begin{eqnarray}
\label{eq:myP}
P(t;p) &=& - \frac{3 |1-t| }{32 p^6 t}
             \left[1 + (10+8p^2+4p^4)t+t^2 \right] \nonumber\\
&& + \frac{3 (1+t)}{8 p^5 } \bigg\{ \frac{3+3p^2+p^4}{\sqrt{1+p^2}}\nonumber\\
&& - \frac{3+2 p^2}{2p} \left[2\, \rmn{arcsinh} (p) - |\ln(t)|\right]\bigg\}. 
\end{eqnarray}
The allowed range of frequency shifts is restricted to
\begin{equation}
|\ln(t)| \le 2\,\rmn{arcsinh}(p)\,,
\end{equation}
and thus $P(t;p)=0$ for $|\ln(t)| > 2\,{\rm arcsinh}(p)$. Similar expressions
for the photon redistribution function using different variables can be found
in the literature \citep{1995ApJ...445...33R, 1998AA...330...90E,
  2000A&A...354L..53S}.

The spectral distortions owing to the relativistic SZ effect can be rewritten
to include a relativistic Comptonization parameter $\tilde{y}$,
\begin{equation}
  \label{eq:rSZ}
  \delta i_\rmn{rel}(x) = [j(x) - i(x)] \tau_\rmn{rel} = \tilde{g}(x) \tilde{y},
\end{equation}
where 
\begin{eqnarray}
  \label{eq:ytilde}
  \tilde{y} &=&  
  \frac{\sigma_\mathrm{T}}{m_\e c^2}\int\dd l\,n_\rmn{e}\, k \tilde{T}_\e\,,\\
  k\tilde{T}_\e &=& \frac{P_\e}{n_\e}\,, \\
  \tilde{g}(x) &=& [j(x) - i(x)]\, \tilde{\beta}(k\tilde{T}_\e)\,,\\
  \label{eq:tildebeta}
  \tilde{\beta}(k\tilde{T}_\e) &=& \frac{m_\e c^2}{\bra k\tilde{T}_\e\ket}
  = \frac{m_\e c^2 \int \dd l\, n_\e}{\int \dd l\, n_\e k\tilde{T}_\e}\,.
\end{eqnarray}
We introduced the normalized pseudo-thermal beta-parameter
$\tilde{\beta}(k\tilde{T}_\e)$ and the pseudo-temperature $k\tilde{T}_\e$ which
are both equal to its thermodynamic analog in the case of a thermal electron
distribution.  If the CRe population is described by the power-law distribution
(\ref{eq:fCRe}), the CRe pressure is given by
\begin{eqnarray}
\label{eq:Pcr}
P_\cre &=& 
\frac{m_\e c^2}{3}\,\int_0^\infty  \!\! \dd p\, f(p)\,\beta_\e\,p  \\
&=& \frac{n_\cre m_\e c^2 (\alpha-1)}{6\left[p^{1-\alpha}\right]_{p_2}^{p_1}} \, 
\left[\B_{\frac{1}{1+p^2}} \left( \frac{\alpha-2}{2},\frac{3-\alpha}{2}
\right)\right]_{p_2}^{p_1} ,
\end{eqnarray}
where $\beta_\e \equiv \upsilon/c = p/\sqrt{1+p^2}$ is the dimensionless velocity
of the electron, $\B_q(a,b)$ denotes the incomplete Beta function
\citep{1965hmfw.book.....A}. In this case, the normalization of the CRe
distribution function $f(p)\dd p$ is determined by the CRe number density,
$n_\cre = \int\dd p f(p)$.  Here, we introduced the abbreviation
\begin{equation}
[F(p)]_{p_2}^{p_1} = F(p_1) - F(p_2)
\end{equation}
in order to account for the lower and upper cutoff $p_1$ and $p_2$ of the CRe
population.

\section{Toy model for plasma bubbles}
\label{sec:toymodel}
This section adopts an analytical formalism to describe buoyant plasma bubbles
which was developed for the analysis of X-ray and radio emission by
\citet{2002A&A...384L..27E}.  After a phase of supersonic propagation of the
radio plasma into the ambient ICM, the radio lobes quickly reach pressure
equilibrium with the surrounding medium once the AGN activity has terminated.
During this stage, the bubble rises with constant velocity governed by the
balance of buoyancy and drag forces while the volume of the bubble expands
adiabatically.  Meanwhile, the surrounding gas is approximately in hydrostatic
equilibrium with the underlying dark matter potential.  Synchrotron, inverse
Compton, and adiabatic losses diminish the observable radio emitting electron
population within the plasma bubble producing a so-called ghost cavity.

As an analytically feasible toy model, we assume spherical geometry
of the plasma bubble and adopt the general $n$-fold $\beta$-profile for the
electron pressure of the ICM which might find application for cool-core
clusters:
\begin{equation}
  \label{eq:pe}
  P_\e(r) = n_\e(r) k T_\e(r) = 
  \sum_{i=1}^N P_i \left[1+\left(\frac{r}{r_{y,i}}\right)^2\right]^{-3\beta_{y,i}/2}.
\end{equation}
The origin of our coordinate system coincides with the cluster center while the
$x_1$- and $x_2$-axes define the image plane, and the $z$-axis the
line-of-sight to the observer. We choose the direction of the $x_1$-axis such
that the bubble center is located in the $x_1$-$z$ plane at $\vec{r}_\rmn{c} =
(r_\rmn{c} \cos\theta, 0, r_\rmn{c}\sin\theta)$. Its projected distance from
the cluster center amounts to $R_\rmn{c}=\mu r_\rmn{c}$ with $\mu =
\cos\theta$, while the radius of the bubble is denoted by $r_\rmn{b}$.  For an
unperturbed line-of-sight which is not intersecting the bubble, the observed
thermal Comptonization parameter $y_\rmn{cl}(x_1,x_2)$ of the cluster is given
by
\begin{equation}
  \label{eq:ycl}
  y_\rmn{cl}(x_1,x_2) = \sum_{i=1}^N y_i 
  \left(1 + \frac{x_1^2+x_2^2}{r_{y,i}^2}\right)^{-(3\beta_{y,i}-1)/2} + y_\rmn{bg},
\end{equation}
where $y_{i} = \sigma_\rmn{T}(m_\e c^2)^{-1} P_i r_{y,i}
\B\left(\frac{3\beta_{y,i}-1}{2},\frac{1}{2}\right)$ is the central thermal
Compton parameter of the respective individual $\beta$-profile and $y_\rmn{bg}$
is the background contribution to the Comptonization which we set to zero in
our analysis. In the case of a line-of-sight intersecting the surface of the
bubble, the two intersection points are $(x_1,x_2,z_\pm)$ with
\begin{equation}
  \label{eq:zpm}
  z_\pm = r_\rmn{c}\sqrt{1-\mu^2} \pm 
  \sqrt{r_\rmn{b}^2 - x_2^2 - (x_1^{}-r_\rmn{c}\mu)^2}.
\end{equation}
The thermal Comptonization parameter $y(x_1,x_2)$ for the area covered by the
bubble is given by
\begin{eqnarray}
  \label{eq:yb}
  y_\rmn{b}(x_1,x_2) &=& y_\rmn{cl}(x_1,x_2) - \sum_{i=1}^N y_i
  \left(1 + \frac{x_1^2+x_2^2}{r_{y,i}^2}\right)^{-(3\beta_{y,i}-1)/2} \nonumber\\
  &&\times\, \left[\frac{\rmn{sgn}(z)}{2}
    \I_{q_{y,i}(z)}
    \left(\frac{1}{2}, \frac{3\beta_{y,i}-1}{2}\right)\right]_{z_-}^{z_+},
\end{eqnarray}
where $\I_q = \B_q(a,b) / \B(a,b)$ denotes the regularized Beta function and
$q_{y,i}(z) \equiv z^2/(r_{y,i}^2 + x_1^2 + x_2^2 + z^2)$.

\begin{table*}[th]
\caption[t]{Parameters of the deprojected electron pressure profiles which obey
  an $n$-fold  $\beta$-profile defined by Eqn.~(\ref{eq:pe}), the
  deprojected electron density profiles (Eqn.~\ref{eq:ne}), and the individual
  parameters of the northern and southern plasma bubbles within the  cool
  core regions of the Perseus cluster and Abell~2052, respectively. The
  parameters $r_\rmn{c}$ and $r_\rmn{b}$  correspond to the distance from the
  cluster center to the bubble center and the bubble radius,
  respectively. The azimuthal angle to the bubble center $\phi$ is measured
  from the axis defined by positive values of the relative right ascension
  while $\eta_\rmn{s}$ denotes the angle of the normal vector of the
  mushroom-shaped southern bubble in A~2052 which we model as a half-sphere.
  In Perseus, $1\arcsec$ corresponds to $0.36\,h_{70}^{-1} \mbox{ kpc}$ and
  in A~2052, $1\arcsec$ corresponds to $0.69\,h_{70}^{-1}  \mbox{ kpc}$. }
\begin{center}
\begin{tabular}{lccccccccc}
\hline 
\hline
& \multicolumn{9}{l}{\em{Cluster pressure profile:}}
\vphantom{\Large A} \\
\vphantom{\large A}%
Cluster & $P_1$ & $P_2$ & $P_3$ & $r_{y,1}$ & $r_{y,2}$ & $r_{y,3}$ & $\beta_{y,1}$ & 
  $\beta_{y,2}$ & $\beta_{y,3}$ \\
& $[h_{70}^{1/2}\mbox{ keV cm}^{-3}]$ & $[h_{70}^{1/2}\mbox{ keV cm}^{-3}]$ &
  $[h_{70}^{1/2}\mbox{ keV cm}^{-3}]$ &
  $[h_{70}^{-1}\mbox{ kpc}]$ & $[h_{70}^{-1}\mbox{ kpc}]$ & $[h_{70}^{-1}\mbox{ kpc}]$ & & &
\vphantom{\huge y} \\
\hline 
\vphantom{\Large A}%
A 426 (Perseus) & 0.118  & ~~0.036 & & 47 & 178 & & 0.94~~ & 0.55~~ & \\
A 2052\dotfill  & 1.034  & $-0.723$ & $-0.258$ & 22.6 & 28.4 & 237 & 0.658 & 0.746 & 128 \\
\hline 
\hline
& \multicolumn{9}{l}{\em{Cluster density profile:}}
\vphantom{\Large A} \\
\vphantom{\large A}%
& $n_1$ & $n_2$ & $r_{w,1}$ & $r_{w,2}$ & $\beta_{w,1}$ & $\beta_{w,2}$ & & & \\
& $[h_{70}^{1/2}\mbox{ cm}^{-3}]$ & $[h_{70}^{1/2}\mbox{ cm}^{-3}]$ &
  $[h_{70}^{-1}\mbox{ kpc}]$ & $[h_{70}^{-1}\mbox{ kpc}]$ & & & & & 
\vphantom{\huge y} \\
\hline 
\vphantom{\Large A}%
A 426 (Perseus) & $4.60\times 10^{-2}$ & $4.8\times 10^{-3}$ 
                & 57 & 200 & 1.2~~ & 0.58 & & & \\
A 2052\dotfill  & $3.75\times 10^{-2}$  & $2.9\times 10^{-3}$ 
                & 29 & 197 & 0.84 & 0.80 & & & \\
\hline 
\hline 
& \multicolumn{4}{l}{\em{Southern bubble:}} 
& \multicolumn{3}{l}{\em{Northern bubble:}} && 
\vphantom{\Large A} \\
\vphantom{\large A}\vphantom{\huge y}%
& $r_\rmn{c,s}$ & $r_\rmn{b,s}$ & $\phi_\rmn{s}$ & $\eta_\rmn{s}$ &
  $r_\rmn{c,n}$ & $r_\rmn{b,n}$ & $\phi_\rmn{n}$  & &  \\
\hline 
\vphantom{\Large A}%
A 426 (Perseus) & $33\arcsec$ & $23\arcsec$ & ~~$-50\degr$ & & 
                  $17\arcsec$ & $17\arcsec$ & $90\degr$ & &  \\
A 2052\dotfill  & $4.4\arcsec$ & $16.4\arcsec$ & $-128\degr$ & $77\degr$ & 
                  $13\arcsec$ & $11\arcsec$ & $90\degr$ & &  \\ 
\hline 
\end{tabular}
\end{center}
\label{tab:parameters}
\end{table*}

The amplitude of the kinetic SZ effect is proportional to the line-of-sight
integrated electron density for which we also assume  a general $n$-fold
$\beta$-profile:
\begin{equation}
  \label{eq:ne}
  n_\e(r) = \sum_{i=1}^N  n_i 
  \left[1+\left(\frac{r}{r_{w,i}}\right)^2\right]^{-3\beta_{w,i}/2}.
\end{equation}
To avoid confusion, we adopt the notation $r_{w,i}$ and $\beta_{w,i}$ for the
usual core radii and the $\beta$-parameters in analogy to the thermal SZ
effect.  For an unperturbed line-of-sight which is not intersecting the bubble,
the observed kinetic Comptonization parameter $w_\rmn{cl}(x_1,x_2)$ of the
cluster is given by
\begin{equation}
  \label{eq:wcl}
  w_\rmn{cl}(x_1,x_2) = \sum_{i=1}^N w_i 
  \left(1 + \frac{x_1^2+x_2^2}{r_{w,i}^2}\right)^{-(3\beta_{w,i}-1)/2},
\end{equation}
where $w_{i} = \bar{\beta}_\gas\sigma_\rmn{T} n_i r_{w,i}
\B\left(\frac{3\beta_{w,i}-1}{2},\frac{1}{2}\right)$ is the central kinetic
Compton parameter of the respective individual $\beta$-profile.  The kinetic
Comptonization parameter $w(x_1,x_2)$ for the area covered by the bubble is
obtained by analogy to the previous case:
\begin{eqnarray}
  \label{eq:wb}
  w_\rmn{b}(x_1,x_2) &=& w_\rmn{cl}(x_1,x_2) - \sum_{i=1}^N w_i
  \left(1 + \frac{x_1^2+x_2^2}{r_{w,i}^2}\right)^{-(3\beta_{w,i}-1)/2} \nonumber\\
  &&\times\, \left[\frac{\rmn{sgn}(z)}{2}
    \I_{q_{w,i}(z)}
    \left(\frac{1}{2}, \frac{3\beta_{w,i}-1}{2}\right)\right]_{z_-}^{z_+}.
\end{eqnarray}
where $q_{w,i}(z) \equiv z^2/(r_{w,i}^2 + x_1^2 + x_2^2 + z^2)$.

\section{Plasma bubbles of Perseus and Abell 2052}
\label{sec:perseus}

Two of the most prominent examples of radio plasma bubbles in nearby galaxy
clusters can be observed within the cool core regions of the \object{Perseus
  cluster} (redshift $z_\rmn{Perseus} = 0.0179$) and \object{Abell~2052}
($z_\rmn{A2052} = 0.0348$).  Their proximity makes both clusters suitable
targets for plasma bubble observations.  Both clusters each host two bubbles
which reflect the relic plasma of a past cycle of jet activity in the cD galaxy
at the cluster center.  At the current stage, the two radio lobes are in
approximate pressure equilibrium with the surrounding medium and rise with a
velocity governed by the balance of buoyancy and drag forces while the volume
of the bubbles expands adiabatically.

As in the previous section, the coordinate origin coincides with the cluster
center and the new direction of the $x_1$-axis points towards positive values
of the relative right ascension.  Assuming spherical symmetry of the plasma
bubbles, their three dimensional position relative to the cluster center is
degenerate because of projection effects. Together with the cluster center, the
center of the bubbles form a plane which we assume to be perpendicular to the
line-of-sight $z$. The azimuthal angle to the bubble center $\phi$ is measured
from the $x_1$-axis.

\subsection{Perseus}
Using deprojected electron density and temperature profiles derived from X-ray
observations \citep{2003ApJ...590..225C}, we obtain a pressure profile by
fitting a double $\beta$-profile according to Eqn.~(\ref{eq:pe}).  In
principle, we want the X-ray pressure profile in the absence of the radio
bubbles. Since the observed X-ray surface brightness profile was derived
assuming spherical symmetry including the region of the bubbles and doesn't
extend into the very center of the cluster due to the AGN at the center and the
{\it XMM/Newton} point spread function, our calculations somewhat underestimate
the SZ effect of the bubbles.  Table~\ref{tab:parameters} shows the individual
parameters of the two plasma bubbles which are measured from the X-ray image of
the central region of Perseus \citep{2000MNRAS.318L..65F}.

In our model, we adopt the peculiar velocity of the Perseus cluster of
$\bar{\upsilon}_\rmn{gas} = -136\mbox{ km s}^{-1}$ with respect to the rest
frame of the CMB and approaching the observer
\citep{1997MNRAS.291..488H}.\footnote{Using the fundamental plane,
  \citet{1997MNRAS.291..488H} measure the Malmquist bias corrected distance to
  early-type galaxies of the Perseus cluster. Comparing these distances to the
  mean of the individual galaxy redshifts, they infer the peculiar velocity of
  the galaxy cluster, $\upsilon_\rmn{pec} = (-136 \pm 307) \mbox{ km s}^{-1}$,
  where the uncertainty derives from the distance error added in quadrature
  with the cluster mean redshift error. On the other hand, the predicted
  peculiar velocity derived from the IRAS redshift survey density field yields
  $\upsilon_\rmn{pec} = +180 \mbox{ km s}^{-1}$ where the velocity field was
  smoothed on a scale of $8\,h^{-1}\mbox{ Mpc}$ which corresponds to the scale
  from which the clusters collapsed.}  The induced kinetic SZ effect gives rise
to a small attenuation of the SZ decrement at our fiducial frequency. However,
it also leads to an interesting effect at the crossover frequency $\nu_\rmn{c}
\simeq 217 \mbox{ GHz}$, and produces an enhancement of the SZ increment at
higher frequencies (cf.{\ }Sect.~\ref{sec:kinSZ}).

\subsection{Abell 2052}

Despite the lower central pressure of Abell~2052 compared to Perseus,
Abell~2052 lies at higher Galactic latitudes. Thus, SZ flux confusion with
Galactic dust emission is negligible in this case, which might be an
observational advantage.

The electron density profile of Abell~2052 is obtained by deprojecting the
X-ray surface brightness profile of \citet{1999ApJ...517..627M} by means of the
deprojection formula given in Appendix~A of \citet{2004A&A...413...17P}.  Since
the X-ray surface brightness is represented by a double $\beta$-model, the
resulting density profile equals the square root of two single $\beta$-profiles
added in quadrature. Refitting this electron density profile to match the
profile defined by Eqn.~\ref{eq:ne} yields the parameters given in
Table~\ref{tab:parameters}.
  
In order to model the temperature profiles $T_\e (r)$ for Abell~2052, we
applied the universal temperature profile for cool core clusters proposed by
\citet{2001MNRAS.328L..37A} to data taken from \citet{2001ApJ...558L..15B},
\begin{equation}
\label{Te}
T_\e (r) = T_0 + (T_1 - T_0)\,
\left[ 1+\left( \frac{r}{r_\mathrm{temp}} \right)^{-\eta_\rmn{temp}}\right]^{-1},
\end{equation}
where $T_0 = 1.31 \mbox{ keV}$, $T_1 = 3.34 \mbox{ keV}$, $r_\rmn{temp} =
20.6~h_{70}^{-1}\mbox{ kpc}$, and $\eta_\rmn{temp} = 4.5$.  This equation
matches the temperature profile well up to radii of $\sim 0.3\,r_\mathrm{vir}$,
which is sufficient for our purposes since we are especially interested in the
core region of Abell~2052. Combining the electron density and temperature
profiles yields the radial variation in the gas pressure, which we represent as
a triple $\beta$-profile (Table~\ref{tab:parameters}); for two of the
components, the normalization is negative.  As noted for Perseus, what we
really need is the pressure profile in the absence of the radio bubbles.  The
observed X-ray surface brightness profile was derived assuming spherical
symmetry including the region of the bubbles.  Probably as a result of this,
the adopted pressure profile has more structure and a lower central pressure
than would be true of the pressure profile in the absence of the radio bubbles.
Thus, our calculations somewhat underestimate the SZ effect of the bubbles in
Abell 2052. Since the southern bubble has a mushroom shape, we decided to model
it using a half-sphere with the line-of-sight grazing the face.

For our simulation, Abell~2052 is taken to be at rest in the CMB rest frame
owing to the large uncertainties of the velocity determination with the
fundamental plane method to measure the Malmquist bias corrected distance to
early-type galaxies \citep{2004MNRAS.352...61H,
  Hudson}.\footnote{\citet{2004MNRAS.352...61H} have only 5 galaxies of
  Abell~2052 in their sample and infer a peculiar cluster velocity of
  $\upsilon_\rmn{pec} = (-494 \pm 1062) \mbox{ km s}^{-1}$ where the error is
  dominated by the error in the distance but also includes redshift
  uncertainties and systematic effects such as extinction. On the other hand,
  the predicted peculiar velocity of Abell~2052 according the IRAS redshift
  survey density field is $\upsilon_\rmn{pec} = +208 \mbox{ km s}^{-1}$.  }

\section{Synthetic observations}
\label{sec:synthetic}

\subsection{Atacama Large Millimeter Array}

\begin{figure}[t]
\begin{center}
{\em\large ALMA: Perseus cluster}
\end{center}
\resizebox{\hsize}{!}{\includegraphics{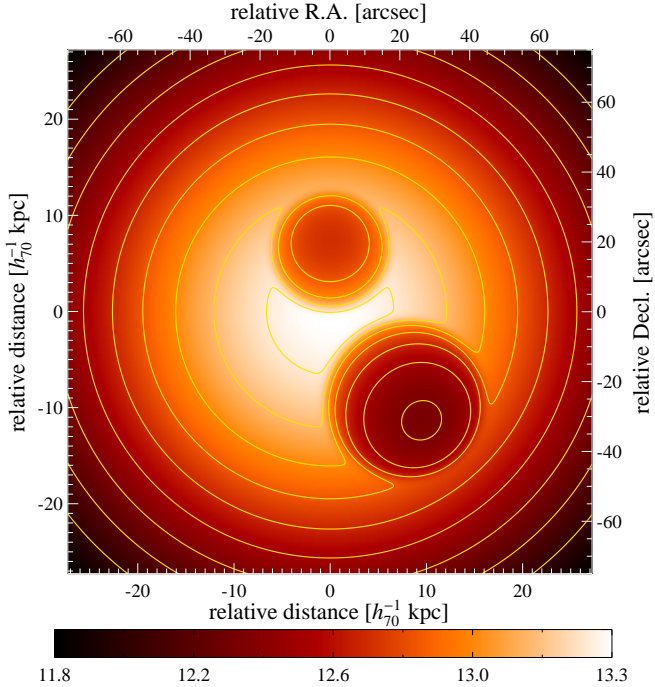}}
\caption{Synthetic {\em ALMA} observation
  of the cool core region of Perseus at the frequency band centered on 144~GHz.
  Shown is the simulated SZ flux decrement of radio plasma bubbles in units of
  $\mbox{mJy/arcmin}^{2}$ assuming an ultra-relativistic electron population
  within the bubbles (scenario 1).  The image is smoothed to the resolution of
  the {\em ALMA} compact core configuration ($\rmn{FWHM} \simeq 2.9\arcsec$).
  The contour lines have a linear spacing of $0.16\mbox{ mJy/arcmin}^{2}$. For
  comparison, the size of the {\em ALMA} field of view at this frequency is
  $\mbox{FWHM}_\rmn{fov} \simeq 36\arcsec$.}
\label{fig:fig1}
\end{figure}

In this section, only the general setup and the results for the synthetic {\em
  ALMA} observation of the cool core regions of Perseus and Abell~2052 are
presented while the interested reader is referred to Appendix~\ref{appendix}
for further details.

To simulate an {\em ALMA} observation of the central region of Perseus, we
compute the frequency band-averaged SZ flux decrement $\bra \delta
I\ket_{\nu_0}$ around the frequency $\nu_0 = 144\mbox{ GHz}$ which samples the
extremum of the SZ flux decrement.  The resulting simulated SZ flux decrement
was convolved with a Gaussian of width $\sigma_{\nu_0} \simeq 1.2\arcsec$ to
obtain the resolution of this {\em ALMA} configuration.  Since Perseus has more
than 40 degrees of declination and is a Northern object, {\em ALMA} will be
challenged to observe it, culminating at 25 degrees elevation.  However, for a
North-South elongated configuration of {\em ALMA} about the factor of $(\cos
\,65\degr)^{-1} \simeq 2.4$, the beam will be almost round.  Assuming an
ultra-relativistic electron population within the bubbles yields the SZ flux
decrement as shown in Fig.~\ref{fig:fig1} which is similar in morphology to the
X-ray image.

To investigate whether the plasma bubbles are detectable in the SZ flux
decrement, we define the SZ flux contrast $\rho$:
\begin{equation}
  \label{eq:contrast}
  \rho = \frac{I_\rmn{A} - I_\rmn{B}}{I_\rmn{A} + I_\rmn{B}}.
\end{equation}
Here, $I_\rmn{A}$ and $I_\rmn{B}$ denote the mean SZ flux decrement of two
equally sized solid angle elements within the field of view, one of which
measures the SZ flux inside and the other one outside the bubble. Prior
information about the angular position of the X-ray cavities allows one to
maximize the SZ flux contrast.  Adopting Gaussian error propagation and
introducing {\em ALMA}'s sensitivity in terms of flux density,
$\sigma_{\!ALMA}$, yields the signal-to-noise for the $5\sigma$ detection of
the bubble:
\begin{equation}
  \label{eq:s2n}
  \frac{\mathcal{S}}{\mathcal{N}} \equiv \frac{\rho}{\sigma_\rho} = 
    \frac{\sqrt{N_\rmn{beam}}\, \left(I_\rmn{A}^2 - I_\rmn{B}^2\right)}
    {2 \sigma_{\!ALMA}\sqrt{I_\rmn{A}^2 + I_\rmn{B}^2}} 
    \ge 5.
\end{equation}
Here, $N_\rmn{beam}$ is the number of statistically independent beams per flux
averaged solid angle element (cf. Appendix~\ref{appendix}).  We choose the
field of view to be centered on the inner rim of the southern bubble such that
equal solid angle elements fall inside and outside the bubble, respectively.
The simulated mean SZ flux decrements within the two solid angle elements of
the primary beam require only an integration time of 5.1 hours in order to
obtain a $5\sigma$ detection of the plasma bubble assuming an
ultra-relativistic electron population within the bubble.  Uncertainties in the
amplitude of the kinetic SZ effect and the geometrical arrangement and shape of
the bubbles may slightly modify this result. As a word of caution, this
observation time allows a single bubble to be observed in a single pointing
while it might be advisable to map the entire central region including both
bubbles to get a clearer picture of the structure, and to be convinced that any
holes seen in the SZ map at the radio bubbles weren't just fluctuations also
seen elsewhere in the cluster center away from the radio bubbles.

The corresponding observation of Abell~2052 would require an integration time
of 38 hours in order to obtain a $5\sigma$ detection of the plasma bubble for
the same plasma bubble content, the longer exposures mainly being the result of
the higher redshift and the lower central temperature of this cluster.

\begin{table}[t]
\caption[t]{Summary of predicted exposure times for the different combinations
  of telescopes and clusters. $I_\rmn{A}$ and $I_\rmn{B}$ denote the mean SZ
  flux decrement of two equally sized solid angle elements within the field of
  view, one of which measures the SZ flux outside and the other one inside the
  bubble. } 
\vspace{-0.1 cm}
\begin{center}
\begin{tabular}{lccc}
\hline \hline
\vphantom{\Large A}%
Telescope: cluster & $I_A$ & $I_B$ & exposure \\
& $[\mbox{mJy amin}^{-2}]$ & $[\mbox{mJy amin}^{-2}]$ & [hours] \\
\hline 
\vphantom{\Large A}%
ALMA: Perseus    & 13.25 & 12.70 & 5.1 \\
ALMA: Abell~2052 & 3.930 & 3.698 & 38  \\
GBT:  Perseus    & 11.31 & 10.85 & 2.1 \\
GBT:  Abell~2052 & 3.272 & 3.138 & 31  \\
\hline 
\end{tabular}
\end{center}
\label{tab:Xcrp}
\end{table}

\subsection{Green Bank Telescope}
\label{sec:GBT}

\begin{figure*}[t]
\begin{tabular}{cc}
{\em\large Green Bank Telescope: Perseus cluster} & 
{\em\large Green Bank Telescope: Abell 2052} \\
\resizebox{0.48\hsize}{!}{\includegraphics{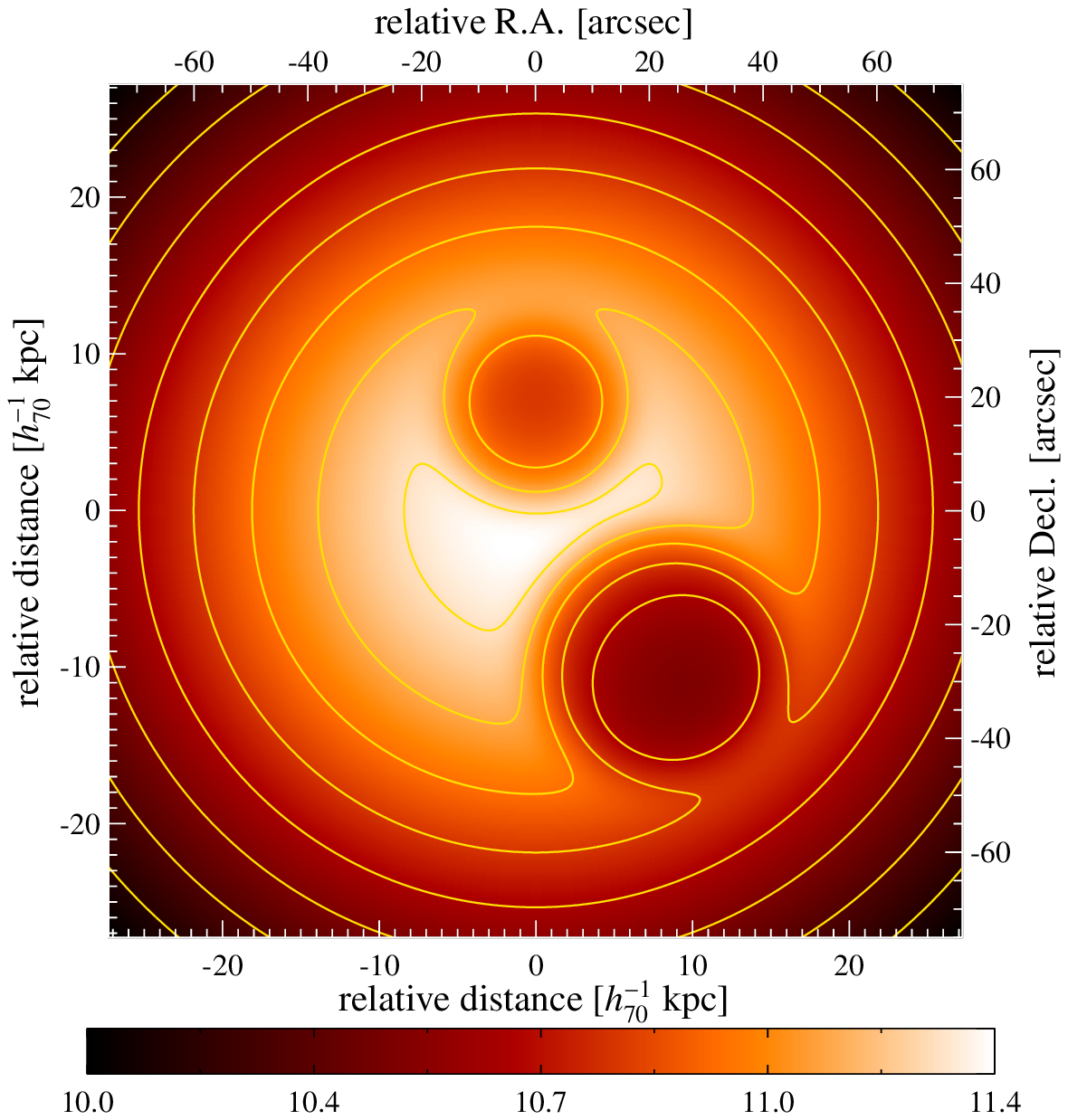}} &
\resizebox{0.48\hsize}{!}{\includegraphics{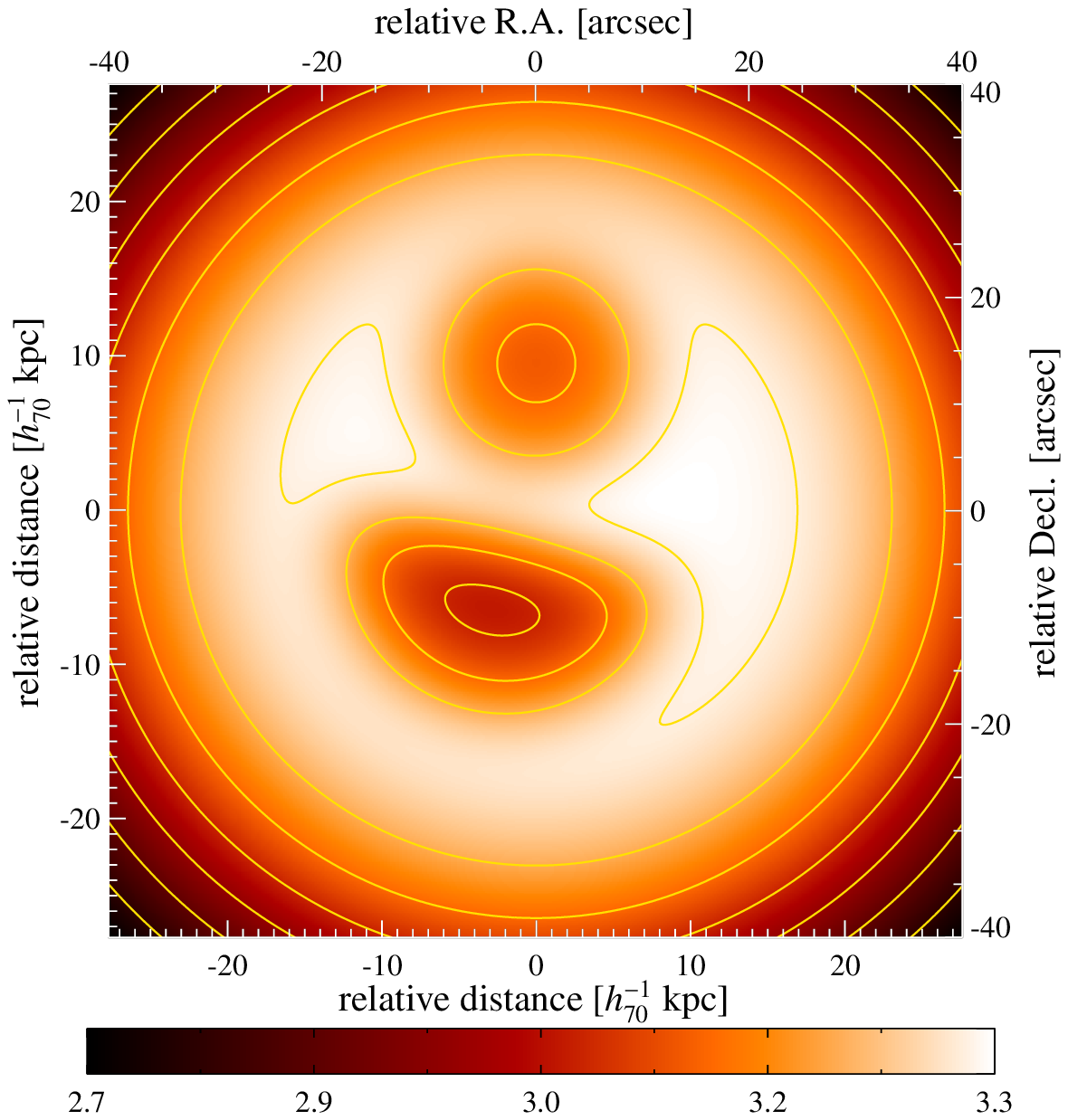}}
\end{tabular}
\caption{Synthetic {\em GBT} observations of the cool core regions of Perseus
    ({\em left panel}) and Abell~2052 ({\em right panel}) at the frequency band
    centered on 90~GHz.  Shown is the simulated SZ flux decrement of radio
    plasma bubbles in units of $\mbox{mJy/arcmin}^{2}$ assuming an
    ultra-relativistic electron population within the bubbles (scenario 1). The
    images are smoothed to the resolution the {\em GBT} 3~mm receiver
    ($\rmn{FWHM} \simeq 8.0\arcsec$).  The contour lines have a linear spacing
    of $0.16\mbox{ mJy/arcmin}^{2}$ ({\em left panel}) and $0.08\mbox{
      mJy/arcmin}^{2}$ ({\em right panel}). For comparison, the size of the
    {\em GBT} field of view at this frequency is $32\arcsec\times 32\arcsec$.}
\label{fig:fig2}
\end{figure*}

To simulate a {\em GBT} observation of the central region of both clusters, we
compute the frequency band-averaged SZ flux decrement $\bra \delta
I\ket_{\nu_0}$ in the frequency interval $[86\mbox{ GHz}, 94\mbox{ GHz}]$.  The
resulting simulated SZ flux decrement was convolved with a Gaussian of width
$\sigma \simeq 3.4\arcsec$ to obtain the resolution of the {\em GBT} 3~mm
receiver.  Assuming an ultra-relativistic electron population within the
bubbles yields the SZ flux decrement as shown in 
Fig.~\ref{fig:fig2}.

To investigate whether the plasma bubbles are detectable in the SZ flux
decrement, we adopt the concept of SZ flux contrast of the previous section.
The sensitivity of the upcoming {\em GBT} Penn Array Receiver in terms of flux
density is given by $\sigma_{\!GBT} = 0.25\, (\Delta t)^{-1/2} \mbox{
  mJy arcmin}^{-2}$, where $\Delta t$ is the integration time in hours
\citep{Mason}.  We choose the $32\arcsec\times 32\arcsec$ field of view of the
{\em GBT} to be centered on the inner rim of the southern bubble of Perseus as
described above. Assuming an ultra-relativistic electron population within the
bubble, we predict a $5\sigma$ detection of the plasma bubble after an
integration time of 2.1 hours owing to the better sensitivity of the bolometric
receivers on the {\em GBT}. This observation time assumes a proper foreground
subtraction of the Galactic emission components.

The corresponding observation of Abell~2052 would require an integration time
of 31~hours in order to obtain a $5\sigma$ detection of the plasma bubble for
the same plasma bubble content. Again, the integration times correspond to a
single pointing on the southern bubble while mapping the entire central region
would take respectively longer.

\section{Composition study of plasma bubbles}
\label{sec:composition}

In the following, we study exemplarily five physically different scenarios of
the composition of the plasma bubble which is as a whole in approximate
pressure equilibrium with the ambient ICM.  Although these scenarios might not
be realized in nature in these pure forms, a realistic SZ flux decrement can be
obtained by linearly combining the different scenarios due to the superposition
property of the pressure of different populations:
\begin{enumerate}
\item The internal pressure is either dominated by CRps, magnetic fields, or
  ultra-relativistic CRes being characterized by a mean momentum of $\bra p\ket
  \gg 1$ in this scenario. This is the most positive case for the detection
  of the plasma bubbles in the SZ flux decrement, as the bubble volume does not
  contribute to the SZ flux decrement significantly. Although this is the
    most positive scenario, it also is the one adopted by most analyses of
    radio lobes.
\item This scenario assumes the internal pressure to be dominated by a compound
  of CRp and CRe populations where the latter is described by a power-law
  distribution:
  \begin{equation}
    \label{eq:fCRe}
    f_\cre (p, \alpha, p_1, p_2) = 
    \frac{(\alpha-1) p^{-\alpha}}{p_1^{1-\alpha}-p_2^{1-\alpha}}.
  \end{equation}
  The distribution function is normalized such that its integral over momentum
  space yields unity. The choice of $p_1=1$, $p_2=10^3$, and $\alpha=2$ implies
  a mean momentum of $\bra p\ket \simeq 6.9$ as well as a pseudo-temperature of
  $k\tilde{T}_\cre \simeq 2.2 \,m_\e c^2 \simeq 1.1 \mbox{ MeV}$ and represents
  a plausible scenario for the relativistic composition of the bubble.  The CRp
  and CRe populations each contribute equally to the internal pressure of the
  bubble representing a remnant plasma originating from the hadronic jet
  scenario.
\item The dynamically dominant internal pressure support is contributed to
  equal amounts by relativistic electron and positron populations,
  respectively. Taking the same parameters for the CRe distribution of the
  previous scenario, this approach represents the remnant radio plasma
  originating from the electron-positron jet scenario.
\item A trans-relativistic thermal proton and electron distribution with $k
  T_\e = 50 \mbox{ keV}$ dominates dynamically over the other non-thermal
  components:
  \begin{equation}
    \label{eq:fth}
    f_\rmn{e,th} (p, \beta_\rmn{th}) = 
    \frac{\beta_\rmn{th}}{K_2(\beta_\rmn{th})} 
    p^2 \exp\left(-\beta_\rmn{th}\sqrt{1+p^2}\,\right).
  \end{equation}
  $K_2$ denotes the modified Bessel function of the second kind
  \citep{1965hmfw.book.....A} which takes care of the proper normalization and
  $\beta_\rmn{th} = m_\e c^2/ (k T_\e)$ is the normalized thermal
  beta-parameter. The mean momentum of this distribution amounts to $\bra
  p\ket \simeq 0.55$.
\item This scenario assumes a dynamically dominant hot thermal proton and
  electron distribution with $k T_\e = 20 \mbox{ keV}$ which exhibits a mean
  momentum of $\bra p\ket \simeq 0.33$.
\end{enumerate}

In all cases, the bubble's sound velocity and thus the barometric scale height
are much higher than the corresponding values of the ambient ICM which
implies a flat pressure distribution within the plasma bubble. This leads
to a reduced internal pressure with respect to the ambient ICM at the inner rim
of the bubble and an excess pressure at the outer rim of the bubble, which is
responsible for the buoyant rise of the bubble in the cluster atmosphere.

\begin{figure}[t]
\resizebox{\hsize}{!}{\includegraphics{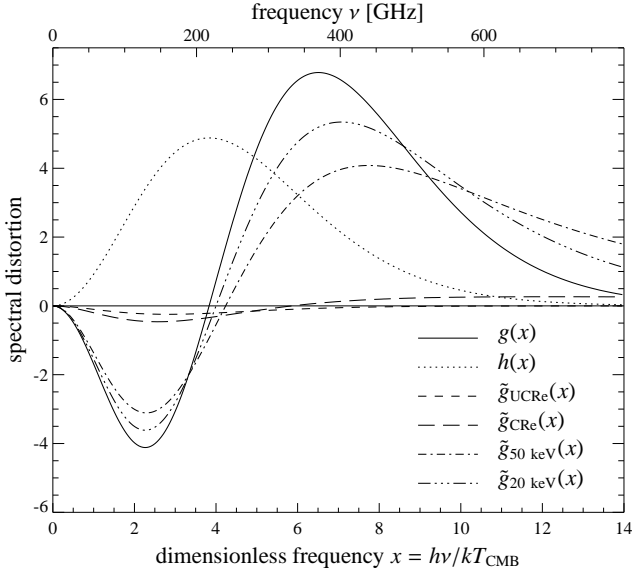}}
\caption{Spectral distortions due to the thermal SZ effect $g(x)$, kinetic SZ
  effect $h(x)$, relativistic SZ effect due to a population of
  ultra-relativistic CRes, $\tilde{g}_\rmn{UCRe} = -i(x)\,
  \tilde{\beta}_\rmn{UCRe}$ (with $p_1=3$, $p_2=10^3$, and $\alpha=2$),
  relativistic SZ effect due to a population of power-law CRes,
  $\tilde{g}_\rmn{CRe} = [j(x)-i(x)]\,\tilde{\beta}_\rmn{CRe}$ (with $p_1=1$,
  $p_2=10^3$, and $\alpha=2$), and the relativistic SZ effect due to a
  population of trans-relativistic thermal electrons, $\tilde{g}_\rmn{50~keV} =
  [j(x)-i(x)]\,\beta_\rmn{th}(\rmn{50~keV})$, as well as due to electrons with
  $k T_\e = 20 \mbox{ keV}$, respectively.}
\label{fig:fig3}
\end{figure}

Figure~\ref{fig:fig3} shows the spectral distortions due to the thermal SZ
effect $g(x)$, kinetic SZ effect $h(x)$, and relativistic SZ effect of the
various scenarios for the relativistic populations of the bubble composition.
The spectral distortions of the relativistic SZ effect are given by
$\tilde{g}(x)$ which have been defined in Eqns.~(\ref{eq:rSZ}) through
(\ref{eq:tildebeta}). Please note, that the scenarios 2 and 3 exhibit the same
spectral distortion $\tilde{g}_\rmn{CRe}$ and that the amplitude of the kinetic
SZ effect $w_\rmn{gas}$ is typically one order of magnitude smaller than the
amplitude of the thermal SZ effect $y_\rmn{gas}$.

\begin{figure*}[ht]
\begin{tabular}{cc}
\resizebox{0.48\hsize}{!}{\includegraphics{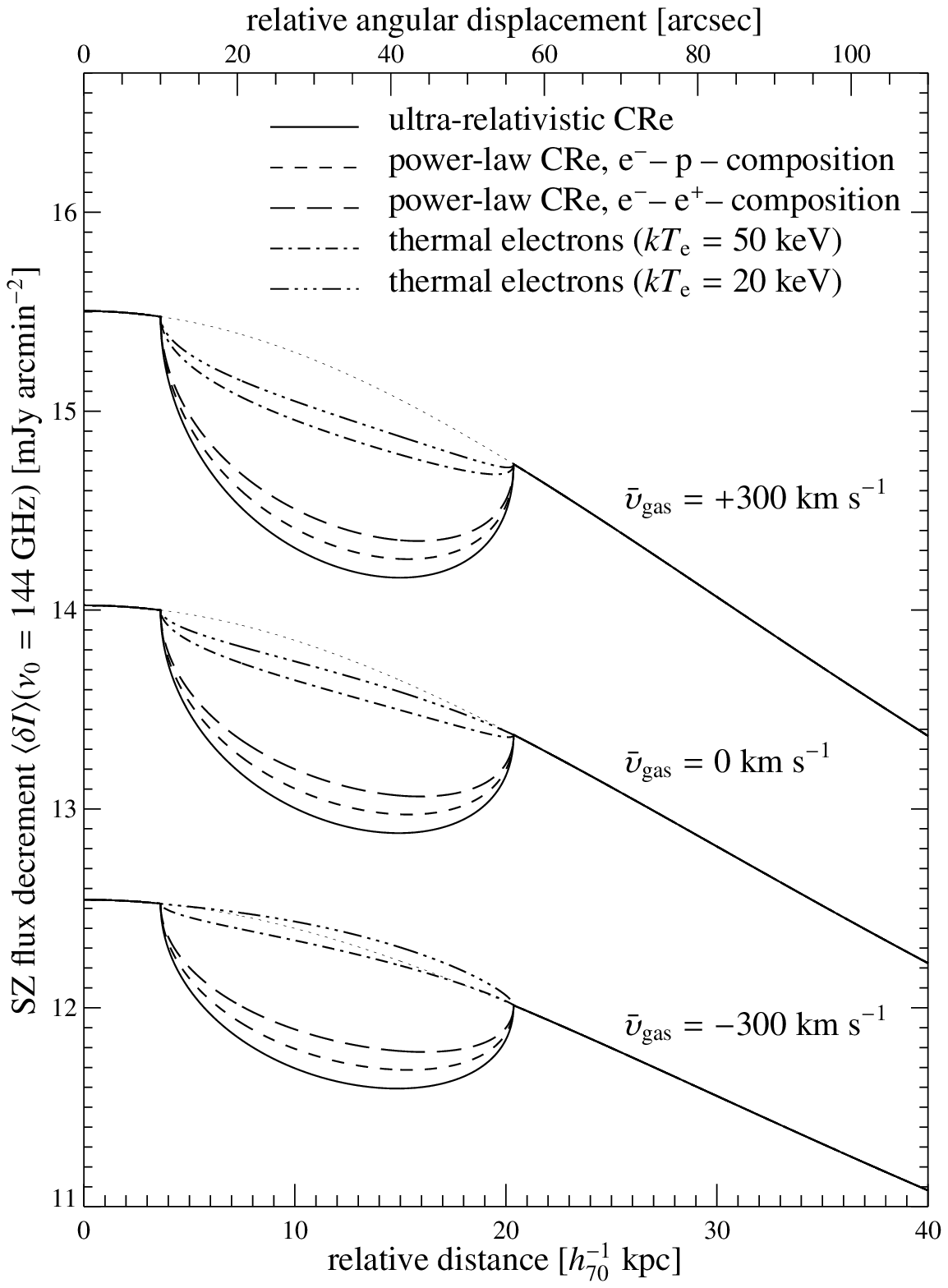}} &
\resizebox{0.48\hsize}{!}{\includegraphics{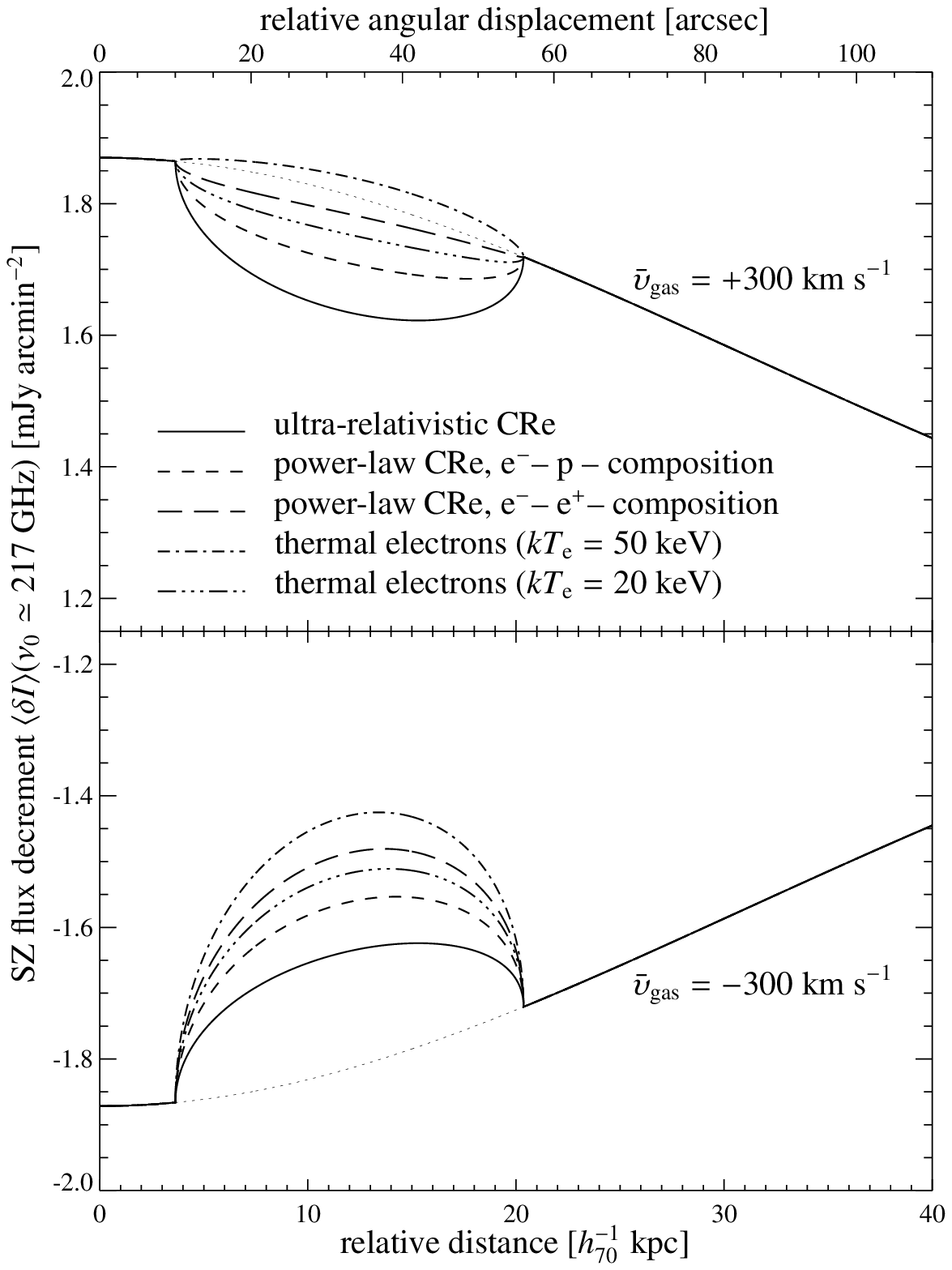}}
\end{tabular}
\caption{Unconvolved SZ flux decrement along an impact parameter through the
  center of the southern bubble of Perseus. The {\em left panel} shows the SZ
  flux decrement at the {\em ALMA} frequency band centered on $\nu_0 = 144
  \mbox{ GHz}$ while the {\em right panel} assumes a central fiducial frequency
  of $\nu_0 \simeq 217 \mbox{ GHz}$.  Compared are five different scenarios of
  the composition of the plasma bubbles to the undisturbed SZ profile (dotted
  thin line), respectively. The three (two) set of lines correspond to three
  (two) differently assumed average bulk velocities along the line-of-sight,
  $\bar{\upsilon}_\gas $, of the thermal gas of Perseus.  }
\label{fig:fig4}
\end{figure*}

The left panel of Fig.~\ref{fig:fig4} shows the unconvolved SZ flux decrement
along an impact parameter through the center of the southern bubble of Perseus
at the {\em ALMA} frequency band centered on $\nu_0 = 144 \mbox{ GHz}$.  While
the northern bubble of Perseus would evince qualitatively the same behavior, it
shows a shallower depth of its SZ cavity due to its smaller geometrical size
resulting in a weaker SZ flux contrast (cf.{\ }Fig.~\ref{fig:fig1}).  The depth
of the SZ cavity at this frequency range is a measure how relativistic the
respective electron population is, i.e.{\ }a deeper SZ cavity indicates a
higher mean momentum of the electron population.  Our studies show a strong
signature of the different bubble compositions on the SZ flux decrement.
Thus, the combination of X-ray and SZ observations allows one to circumvent
  the degeneracy between the effects of the bubble composition and of the
  bubble extent along the line-of-sight on the SZ measurement.   This enables
us to distinguish a relativistic from a thermal electron population inside the
bubble using only a single frequency SZ observation by either a detection or
non-detection of the bubble, respectively.

\section{Kinetic Sunyaev-Zel'dovich effect}
\label{sec:kinSZ}

\subsection{General considerations}
If the cluster is moving towards the observer, the CMB temperature in this
direction is increased due to the Doppler effect of the bulk motion of the
cluster relative to the rest frame of the CMB (in this case, our convention is
such that $\bar{\upsilon}_\rmn{gas}<0$).  This enhanced SZ emission implies a
reduced SZ flux decrement for frequencies below the crossover frequency of the
thermal SZ effect at $\nu_\rmn{c}\simeq 217 \mbox{ GHz}$ (for the
non-relativistic case).  Cosmologically, line-of-sight cluster velocities
follow a Gaussian distribution with vanishing mean since the large scale
structure is at rest in the comoving CMB-frame and with a standard
deviation of $\sigma_\rmn{vel} \simeq 310 \mbox{ km s}^{-1}$ as derived from a
cosmological structure formation simulation comprising the Hubble volume
\citep{2001MNRAS.321..372J, astro-ph/0407089}.

Theoretically, the turbulent motions of the ICM should also contribute to the
kinetic SZ effect. The largest impact is caused by two merging clusters along
the line-of-sight which induces a bimodal streaming flow pattern on the sky.
However, radio plasma bubbles are mainly observed within cool-core clusters
which represent relaxed clusters where the ICM is approximately in hydrostatic
equilibrium with the underlying dark matter potential.  In this case, only
small scale turbulent vortices are expected which exhibit smaller angular
momenta. For an isotropic distribution of vortex orientations, line-of-sight
integration and beam convolution average the resulting kinetic SZ effect out to
zero. Thus, these turbulent motions can only contribute in second order to the
thermal SZ effect.

It would be useful to know the peculiar velocity of the galaxy cluster in
order to remove the degeneracy of the SZ cavity depth with respect to the
different bubble compositions and the kinetic SZ effect. In the case of a
general morphology of a galaxy cluster, in particular for non-axis-symmetric
objects, it is impossible to unambiguously deproject the cluster in order to
  derive the peculiar velocity given only a single frequency SZ observation
  and an X-ray image \citep{1998ApJ...500L..87Z, 2001ApJ...561..600Z}. For
multi-frequency SZ observations of the entire cluster,
\citet{2003JCAP...05..007A} theoretically discuss possibilities in order to
break parameter degeneracies between the Compton parameter, the electron
temperature and the cluster peculiar velocity for an appropriate choice of
observing frequencies. Considering possible gaseous substructure along the
line-of-sight towards the radio plasma bubble and a possibly non-spherical
bubble geometry, it might be impossible to discriminate between morphologically
similar bubble compositions (scenarios 4 and 5) in the case of a single
frequency {\em ALMA} or {\em GBT} observation.

\subsection{Perseus plasma bubbles at different frequencies}
The left panel of Fig.~\ref{fig:fig4} compares three set of lines corresponding
to different average line-of-sight streaming velocities at the {\em ALMA}
frequency band which samples the extremum of the thermal SZ flux decrement
($\nu_0 = 144 \mbox{ GHz}$).  The effect of an overall decrease in the SZ flux
decrement for the approaching cluster ($\bar{\upsilon}_\rmn{gas} = -300 \mbox{
  km s}^{-1}$) can be clearly seen.  The amplitude of the kinetic SZ effect
depends on the dimensionless average velocity of the thermal gas, which amounts
in our case to $\bar{\beta}_\gas = \bar{\upsilon}_\gas / c \simeq 10^{-3}$.
On the other hand, the amplitude of the thermal SZ effect depends on the
inverse normalized thermal beta-parameter, $\beta_\rmn{th}^{-1} = k T_\e/ (m_\e
c^2)\simeq 10^{-2}$, where we inserted the average temperature of the Perseus
cluster. At the extremum of the thermal SZ flux decrement, the kinetic SZ flux
amounts to an approximately $10\%$ correction to the thermal SZ effect for the
choice of our fiducial cluster velocity. However, this small effect is
responsible for the qualitative difference of the observable SZ cavities
resulting from plasma bubbles: assuming a receding cluster and a dynamically
dominant thermal proton and electron distribution (scenarios 4 and 5), we still
obtain a detectable depth of the SZ cavity which almost disappears for an
approaching cluster.

The right panel of Fig.~\ref{fig:fig4} shows the unconvolved SZ flux decrement
assuming a central frequency of $\nu_0 = 217.34 \mbox{ GHz}$ and bandwidth of
$\Delta \nu = 40 \mbox{ GHz}$.  Allowing for finite frequency response of the
instrument's receivers, this fiducial frequency corresponds to a vanishing
frequency band-averaged thermal SZ flux decrement. Thus, the kinetic SZ
effect represents the main effect at this frequency showing a positive SZ flux
decrement for a receding cluster, i.e.{\ }we would detect a reduced flux of
CMB photons in that direction.  At this frequency range, a relativistic electron
population always causes a positive relativistic SZ flux decrement owing to the
higher crossover frequency $\nu_\rmn{c} \gtrsim 217 \mbox{ GHz}$ and
irrespective of the cluster's velocity. This effect causes an enhanced SZ flux
contrast of the bubble for an approaching cluster compared to a receding one,
depending on the specific composition of the plasma bubble. 

Most remarkable is the interchange of SZ flux decrements for different bubble
compositions compared to the case of $\nu_0 = 144 \mbox{ GHz}$.  The largest
impact of the bubble's SZ cavity is provided by the trans-relativistic thermal
electron population of 50~keV owing to its comparably large frequency
band-averaged SZ flux decrement. The SZ flux decrements of the other bubble
compositions are smaller because the crossover frequency of the thermal
electron population of $k T_\e = 20 \mbox{ keV}$ lies closer to the
non-relativistic crossover frequency while the amplitudes of the spectral
distortions of the relativistic electron populations are smaller (cf.{\ 
  }Fig.~\ref{fig:fig3}).  Relativistic corrections to the thermal SZ flux
resulting from the thermal electrons with temperatures of $k T_\e \simeq 3-7
\mbox{ keV}$ are negligible: The resulting profiles of the SZ flux decrement
are similar in morphology to the kinetic SZ flux decrement and would correspond
at this frequency to an additional kinetic SZ effect of only $\sim 10 \mbox{ km
  s}^{-1}$.

\begin{figure}[t]
\resizebox{\hsize}{!}{\includegraphics{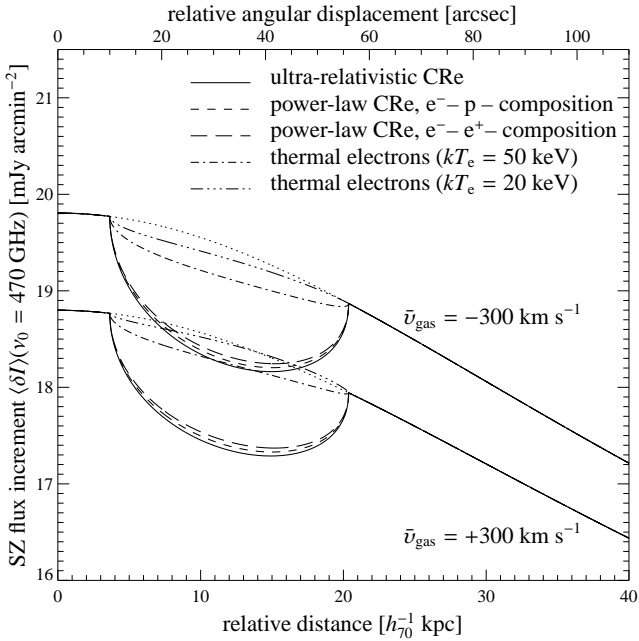}}
\caption{Unconvolved SZ flux increment along an impact parameter through the
  center of the southern bubble of Perseus at the frequency band centered on
  $\nu_0 \simeq 470 \mbox{ GHz}$. Compared are five different scenarios of the
  composition of the plasma bubbles to the undisturbed SZ profile (dotted thin
  line), respectively. The two set of lines correspond to two differently
  assumed average bulk velocities along the line-of-sight,
  $\bar{\upsilon}_\gas$, of the thermal gas of Perseus.  }
\label{fig:fig5}
\end{figure}

Owing to the proximity of the Perseus cluster to the Galactic plane, SZ
observations at even higher frequencies might be challenging due to the
Galactic dust emission which represents the major Galactic foreground at these
frequencies \citep[see e.g.][]{1998ApJ...500..525S, 1999ApJ...524..867F,
  2000ApJ...544...81F, astro-ph/0407090}.  In an exemplary manner, we also show
the expected SZ cavity around a fiducial frequency which samples the maximum of
the SZ flux increment.\footnote{This might find application for a galaxy
  cluster with an X-ray cavity at high Galactic latitude (e.g., Abell~2052)
  or in the case of Perseus, once the Galactic dust emission is properly
  removed.}  For a cluster with a temperature of 5~keV, the normal thermal
relativistic corrections disappear at $\nu_0 = 470 \mbox{ GHz}$ when taking
into account the detector's finite frequency response of $\Delta \nu = 40
\mbox{ GHz}$.  Secondly, at this frequency the kinetic SZ effect has only a
small additional contribution \citep{2003JCAP...05..007A}. Hence, one could
expect that the SZ signal from the bubble could be even more prominent. This is
investigated in Fig.~\ref{fig:fig5} which shows the unconvolved SZ flux
increment along an impact parameter through the center of the southern bubble
of Perseus at $\nu_0 = 470 \mbox{ GHz}$. The different bubble compositions
qualitatively show the same behavior as for the SZ decrement at the frequency
band around $\nu_0 = 144 \mbox{ GHz}$. However, the SZ flux contrast of the
bubble is much higher as well as the degeneracy owing to the kinetic SZ effect
is reduced. Thus, for clusters not being outshone by the Galactic dust
emission, observations at this frequency range seem to be most promising in
order to infer the bubble composition.

\section{Observing strategy}
\label{sec:strategy}

The previous results imply the following observing strategy for the {\em ALMA}
compact core configuration or the {\em GBT}. A short duration observation
(a few hours) of the inner rim of the southern X-ray cavity within the
Perseus cool-core region should either yield a $5\sigma$ detection of an SZ
cavity or not. The case of a significant detection either proves the existence
of a dynamically dominating CRp population, magnetic fields, or an
ultra-relativistic CRe population within the radio plasma bubble while the
contrary excludes this hypothesis on the $5\sigma$ level.

Such a non-detection of the bubble would require a longer integration time in
order to detect the SZ flux decrement of the plasma bubble at the desired
significance level. A detection of a very shallow SZ cavity indicates a
dynamically dominant hot thermal electron population within radio plasma
bubbles.  The underlying physical scenario is degenerate for an observed
intermediate depth of the SZ cavity (such as an SZ flux level in between our
scenarios 3 and 4).  Possibilities include trans-relativistic thermal electron
distributions or soft CRe power-law distributions exhibiting a spectral index
of approximately $\alpha=3$ as well as a small lower cutoff $p_1$ implying a
mean momentum of $\bra p\ket \sim 1$.  Follow-up multi-frequency SZ
observations in combination with X-ray spectroscopy could disentangle the
different scenarios and possibly estimate the temperature or spectral
characteristics of the dynamically dominating electron population
\citep{2002ApJ...575...12S, 2004astro.ph..1337E}.  The detection of such an
  SZ flux would enable one to draw conclusions concerning the particular jet
scenario being responsible for the inflation of the plasma bubble.

The case of a deep SZ cavity with an associated detection of dynamically
dominating CRps, magnetic fields, or ultra-relativistic CRes leads to an
immediate question about the composition of radio ghosts. Thus, an additional
SZ observation of the ghost cavity in Perseus could yield answers about the
potential entrainment of ICM into the plasma bubble during its buoyant rise in
the cluster atmosphere. If a large fraction of entrained gas in the bubble
provides significant pressure support, it must have experienced Coulomb heating
through CRes in order not to be detected in X-rays. This would result in a
reduced SZ flux contrast of the ghost cavity and leads to a faint or
undetectable SZ flux decrement of the ghost cavity.

\section{Conclusion and outlook}
This work provides a theoretical framework for studying the SZ decrement of
radio plasma bubbles within clusters of galaxies. X-ray observations are
proportional to the square of the thermal electron density and probe the core
region of the cluster. On the other hand, the SZ effect is proportional to the
thermal electron pressure enabling the detection of plasma bubbles further
outwards of the respective clusters.  Assuming spherically symmetric plasma
bubbles, we simulate an {\em ALMA} and a {\em GBT} observation of the cool core
regions of the Perseus cluster and Abell~2052. In this context, we exemplarily
investigate physically different scenarios of the composition of the plasma
bubbles: As long as the bubble is dynamically dominated by relativistic
electrons, protons, or magnetic fields, there exists a realistic chance to
detect plasma bubbles in the SZ flux decrement. Non-detection of radio bubbles
with the SZ effect at the position of X-ray cavities hints towards a
dynamically dominant hot thermal electron population within radio plasma
bubbles.  Detection of a non-thermal pressure support should be possible within
a few hours observation with ALMA or GBT in the case of Perseus and a few ten
hours in the case of Abell~2052.

Pursuing high-sensitivity multi-frequency SZ observations, it will be
challenging but not impossible to infer the detailed nature of the different
possible populations of the bubble.  For realistic observations, the frequency
dependence of the relativistic SZ signal is contaminated by the presence of the
kinetic SZ signal.  It would be optimal to have a frequency channel centered
on the crossover frequency of the thermal SZ effect in order to infer the
energy spectrum of the dynamically dominating population and to measure the
temperature or spectral characteristics of the electron population.

\begin{acknowledgements}
  The authors would like to thank Bj{\"o}rn Malte Sch{\"a}fer, Matthias
  Bartelmann, Eugene Churazov, our editor Francoise Combes and an anonymous
  referee for carefully reading the manuscript and their constructive remarks.
  Furthermore, we thank Brian Mason and Jim Condon for discussions about the
  sensitivity of the PAR receiver on the GBT and Mike Hudson for providing the
  peculiar cluster velocities.
\end{acknowledgements}

\appendix

\section{Synthetic {\em ALMA} observation of the central region of Perseus}
\label{appendix}

This appendix provides additional details of the synthetic {\em ALMA}
observation of the cool core region of Perseus as well as the expected SZ flux
contrast of the southern X-ray cavity.  First, we compute the frequency
band-averaged SZ flux decrement which is defined by
\begin{equation}
  \label{app:band_average}
  \bra \delta I\ket_{\nu_0} = 
  \frac{\int\dd\nu\, \delta I(\nu) R_{\nu_0}(\nu)}
  {\int\dd\nu\, R_{\nu_0}(\nu)}.
\end{equation}
Here, $ R_{\nu_0}(\nu)$ denotes the frequency response of the {\em ALMA} receivers
centered on a fiducial frequency $\nu_0$ which we assume to be described by a
top-hat function:
\begin{equation}
R_{\nu_0}(\nu) = 
\left\{
\begin{array}{l@{,\:}l}
1  & \nu\in   \left[\nu_0-\Delta\nu/2,\nu_0+\Delta\nu/2\right] \\
0  & \nu\notin\left[\nu_0-\Delta\nu/2,\nu_0+\Delta\nu/2\right] 
\end{array}
\right.
\label{app:freq_resp}
\end{equation}
We choose the {\em ALMA} frequency band 4 which samples the extremum of the SZ
flux decrement and is characterized by $\nu_0 = 144\mbox{ GHz}$ and $\Delta\nu
= 38\mbox{ GHz}$ \citep{ALMAProjectBook}. The requirement of obtaining the
highest flux sensitivity to the largest scales comparable to the field of view
at this frequency ($\mbox{FWHM}_\rmn{fov} \simeq 36\arcsec$) calls for the most
compact configuration {\em ALMA~E} with a maximal baseline of $d_\rmn{bl} =
150\mbox{ m}$. Thus, we convolve the simulated SZ flux decrement with a
Gaussian to obtain the resolution of this configuration, $\mbox{FWHM} \simeq
c/(\nu \,d_\rmn{bl})\simeq 2.86\arcsec$.  Assuming an ultra-relativistic
electron population within the bubbles yields the SZ flux decrement as shown in
Fig.~\ref{fig:fig1}. 

To investigate, whether the plasma bubbles are detectable in the SZ flux
decrement, we define the SZ flux contrast $\rho$:
\begin{equation}
  \label{app:contrast}
  \rho = \frac{I_\rmn{A} - I_\rmn{B}}{I_\rmn{A} + I_\rmn{B}}.
\end{equation}
Here, $I_\rmn{A}$ and $I_\rmn{B}$ denote the mean SZ flux decrement of two
equally sized solid angle elements within the field of view, one of which
measures the SZ flux inside and the other one outside the bubble. Adopting
  Gaussian error propagation and introducing {\em ALMA}'s sensitivity per beam
  in terms of flux density, $\sigma_{\!ALMA}$, yields the uncertainty in
  $\rho$,
\begin{eqnarray}
  \sigma_\rho
  &=& \sqrt{\left(\frac{\partial \rho}{\partial I_\rmn{A}}\right)^2 +
            \left(\frac{\partial \rho}{\partial I_\rmn{B}}\right)^2} 
      \,\frac{\sigma_{\!ALMA}} {\sqrt{N_\rmn{beam}}} \\
  \label{app:sigma_rho}
  &=& \frac{2 \sqrt{I_\rmn{A}^2 + I_\rmn{B}^2}}
       {\left(I_\rmn{A} + I_\rmn{B}\right)^2}
       \,\frac{\sigma_{\!ALMA}} {\sqrt{N_\rmn{beam}}} .
\end{eqnarray}
Here, $N_\rmn{beam}= f\, A_\rmn{beam}/(2 A_\rmn{fov})$ is the number of
statistically independent beams per flux averaged solid angle element,
$f\le 0.5$ measures the fraction of the solid angle of the bubble within the
  field of view, and the factor of $A_\rmn{beam}/(2 A_\rmn{fov})$ accounts for
  the statistically independent degrees of freedom within the field of view.
Combining Eqns.~\ref{app:contrast} and \ref{app:sigma_rho} yields the
signal-to-noise for the $5\sigma$ detection of the bubble:
\begin{equation}
  \label{app:s2n}
  \frac{\mathcal{S}}{\mathcal{N}} \equiv \frac{\rho}{\sigma_\rho} = 
    \frac{\sqrt{N_\rmn{beam}}\, \left(I_\rmn{A}^2 - I_\rmn{B}^2\right)}
    {2 \sigma_{\!ALMA}\sqrt{I_\rmn{A}^2 + I_\rmn{B}^2}}\ge 5.
\end{equation}
The flux density sensitivity $\sigma_{\!ALMA}$ for point sources, which should
be approximately applicable in the case of the compact core configuration, is
given by \citet{ALMAMemo243}:
\begin{equation}
  \label{app:sigmaA}
  \sigma_{\!ALMA} = 
  \frac{\sqrt{2}\, k T_\rmn{sys}}
  {\eta\, A_\rmn{dish}\, A_\rmn{beam} \sqrt{\Delta t\, \Delta \nu\, N_\rmn{bl}}}.
\end{equation}
Here, $T_\rmn{sys}\simeq 65\mbox{ K}$ denotes the system temperature,
$A_\rmn{dish} = \pi D^2/4$ the collecting area of each dish ($D = 12\mbox{
  m}$), $\eta = 0.75$ the aperture efficiency, $A_\rmn{beam} = 1.13 \mbox{
  FWHM}^2 \simeq 9.3~\rmn{arcsec}^2$ the area of the secondary beam, $\Delta t$
the integration time, $\Delta \nu = 38 \mbox{ GHz}$ the bandwidth, and $
N_\rmn{bl} = n(n-1)/2$ the number of baselines, $n = 64$ being the number of
antennas. We choose the field of view to be centered on the inner rim of the
southern bubble such that equal solid angle elements fall inside and outside
the bubble, respectively. The simulated mean SZ flux decrements within the two
solid angle elements of the primary beam, $I_\rmn{A} = 13.25 \mbox{ mJy
  arcmin}^{-2}$ and $I_\rmn{B} = 12.70 \mbox{ mJy arcmin}^{-2}$, require only
an integration time of 5.1 hours in order to obtain a $5\sigma$ detection of
the plasma bubble assuming an ultra-relativistic electron population within the
bubble. Uncertainties in the amplitude of the kinetic SZ effect and the
geometrical arrangement and shape of the bubbles may slightly modify this
result.

\bibliography{bibtex/chp} \bibliographystyle{bibtex/aa}
\end{document}